# Algorithm stickiness and the memory extent delimit the rationality of El Farol attendees

**Canan Atilgan and Ali Rana Atilgan***

*School of Engineering and Natural Sciences, Sabanci University, Tuzla 34956 Istanbul, Turkey*

**\*Corresponding author:**

atilgan@sabanciuniv.edu
telephone: +90 (216) 483 9525
telefax: +90 (216) 483 9550

**Abstract**

Arthur's paradigm of the El Farol bar for modeling bounded rationality and inductive behavior is undertaken. The memory horizon available to the agents and the selection criteria they utilize for the prediction algorithm are the two essential variables identified to represent the heterogeneity of agent strategies. The latter is enriched by including various rewarding schemes during decision making. Though the external input of comfort level is not explicitly coded in the algorithm pool, it contributes to each agent's decision process. Playing with the essential variables, one can maneuver the overall outcome between the comfort level and the endogenously identified limiting state. Furthermore, we model the behavior of the agents through the use of an expression that scores the local attendance states available to the agents. It incorporates a single parameter that weighs the relative contributions that originate from the external and internal limiting factors. Solving this expression analytically as well as numerically using the Metropolis Monte-Carlo technique enables us to attribute statistical-thermodynamical meanings to the essential variables identified with the agent-based model, and to gain physical insight to the bounds observed in the behavior of the agents. The power of the analytical approach is validated by obtaining a one-to-one correspondence between the agent-based model and the analytical approach within a wide range of thresholds using a single parameter for a given selection criterion. The origin of discrepancies between the two models appearing at extreme thresholds is tracked to the shifts in the distributions of algorithm types actively utilized by the adaptive agents.

**Introduction**

Data is ubiquitously available to each of us. We have been inductively forming our beliefs and perceptions, and adaptively building our expectation formation patterns by observing the past and present events taking place around us. The interplay between what is deduced from the aggregated average behavior and instantaneous forecasts of individuals based upon what is accessible to them shapes the evolutionary path of society's trend.

Inductive methodologies allow agents to construct their own decision making mechanisms from information allocated to all in common. Being an example of this approach, agent-based methodology adopts simple behavioral rules and allows a coordinated equilibrium to be an emergent property of the system, instead of modeling the system as if everyone's actions and beliefs were synchronized a priori with everyone else's (1, 2).

Arthur's El Farol bar problem brings up a paradigm (3). A total number of $N$ players must decide independently whether to attend the bar or not. If a player forecasts that the total attendance will exceed the comfort level, $L$, she will not show up, otherwise she will go. We are interested in the way the players predict the total attendance and its long-term characteristics. The problem is based on the agents' knowledge of the overall attendance history, but the individual actions are not known. Yet, a collective behavior emerges from the system, whereby the average attendance is bounded on two sides by the threshold imposed from the outside, and the randomness that would take over the system in the absence of the threshold.

The problem has been investigated from many perspectives (4-6). In particular, Johnson and coworkers have investigated the variation of volatility with the pool sizes available to the agents, as well as the effect of various selection schemes on attendance (7). Challet et al. have proposed a statistical mechanical model on the problem that becomes exact in the limit of randomness (8). It has also been analytically shown that the problem is mean reverting in nature; therefore, the average attendance will remain within the region [$N/2$, $L$] (9). Yet, a controlled study that explores the effect of the parameters that may influence the output of the system has not been performed. Amongst these are (i) the different types of algorithms utilized by the agents, (ii) the strategy employed to select algorithms from this pool, (iii) the memory horizon for which attendance data is available to the agents, and (iv) the comfort level that is imposed on the system by the outside world. In this study, we systematically study the effect of each of these parameters on the average attendance recorded. We find that the emergent behavior in these systems is not only limited by an upper and lower bound, but also that it follows predictable patterns within these limitations once a critical value of short memory is surpassed. We have further explained the observed patterns using a deductive model on the system that incorporates the main ingredients that affect the outcome.

**Agent-based modeling of the El Farol bar:**

We use a system of $N$ potential attendants. Each week we seek to find the number of people who will actually attend the bar, $a_o$. The attendance data of the last $m$ weeks [$a_m$, $a_{m-1}$, …, $a_1$] are available to the agents. Each attendant, $j$, takes a decision to attend or not by making a prediction, $q_j$, on the possible value of $a_o$, using one of the many available algorithms. The algorithm pool here consists of (i) point-wise hypothesis – the agent uses the attendance data of the $k^{th}$ previous week ($1 \leq k \leq$ *m*) as his prediction, $q_j = a_k$; (ii) arithmetic average – the agent uses the average of the last *k* (1 < *k* ≤ *m*) weeks as his prediction, $q_j = \frac{1}{k}\sum_{i=1,k} a_i$ ; (iii) weighted-average – the agent uses a weighted average of the last *k* (1 < *k* ≤ *m*) weeks, where the more recent a week's data is, the larger weight it has, $q_j = \sum_{i=1,k} \frac{2ia_i}{k(k+1)}$; (iv) trend – the agent makes a least squares fit to the last *k* weeks' data (1 < *k* ≤ *m*), and uses its extrapolation to the following week as his prediction, bounded by [0,100]. There are, therefore, a total of 4*m* – 3 algorithms available, where *m* is the memory of the system. Once the agent makes his prediction $q_j$, he will not attend if the prediction is larger than a previously set comfort level, *L*:

$$h_j = \begin{cases} 1 & q_j \leq L \\ 0 & q_j > L \end{cases} \quad (1)$$

The attendance is then the sum,

$$a_o = \sum_j h_j \quad (2)$$

The algorithm pool available to each agent affects the results. In one scenario, all algorithms in the pool may be available to all agents. However, note that alternatively, a sub-pool specific to each agent may be assigned *a priori*, and the agents may only choose from within this subset. Johnson and coworkers have shown that this choice significantly affects the volatility, where a minimum value of the volatility is obtained for a given fraction of the pool size, and it has a larger value for lower or higher fractions; maximum volatility is recorded for a fraction of one (7). The sub-pool may consist of a particular type of strategy (i) – (iv) above, or a mixture of those. The union of all the sub-pools should then define the total of the algorithms. The algorithms are initially assigned to the sub-pools using a uniform distribution.

The extent to which the memory of each agent extends into the past is equivalent to *m* in all these cases. In what follows, one property we investigate in detail is how *m* affects the outcome. Since the fraction that gives the minimum volatility depends on *m*, we choose to make all the algorithms available to all the agents; i.e. the sub-pool fraction is one. Volatility is therefore at its maximum for the entire memory horizon.

Another property that significantly influences the results is the way the agents pick future algorithms from their sub-pool. This may be done in one of a multitude of ways: In the simplest of the applications, each agent hangs on to his current algorithm as long as it is successful, and changes it as soon as he fails in his prediction (denoted by ***scheme I*** here). Alternatively, various rewarding schemes that evaluate the success of the algorithms in retrospect may be adopted. In one scheme, the agent re-evaluates the predictions of all of his algorithms, and picks the one that would have been successful *and* provides the closest value to $a_o$ (***scheme II***). In another scheme, the agent keeps a log of the success of his algorithms, by giving a point to all algorithms that would have succeeded at the current step. The agent then picks the one that has the highest cumulative score as his next predictor, or randomly selects between algorithms in case of equivalent scores (*scheme III*).

Obviously, we are not limited by the selection schemes listed above and employed in this study. There are other ways of implementing behavior at this level. One example is insistence, where the agent changes his current algorithm only after a predetermined number of successive failures. Rewarding may also be implemented in alternative ways, such as an intermediate case of schemes II and III, where the cumulative score is kept on only the best predictor, or by having a memory effect on the cumulative score, etc. As we shall see, such a choice significantly affects the fundamental properties of the system.

**Effect of memory on mean attendance and distributions of active algorithms**

In figure 1 we display the variation of the average attendance with increasing memory for the three schemes discussed above. We observe that there are upper and lower bounds on the attendance: As the amount of information available to the agents increases, the average attendance approaches a value that is equal to or less than the comfort level, *L*. The ultimate value attained, $\gamma$, depends on the choice of the algorithm selection scheme, I, II, or III. The lower bound, on the other hand, tends to *N*/2; i.e. when there is very limited amount of information provided to the agents, their predictions become randomized, irrespective of the comfort level. The change from the lower bound, *N*/2, to the upper bound, $\gamma$, occurs in an S-shaped curve; however, for short memories, depending on the selection scheme used, some of the data points deviate from this curve, as shown with the hollow circles in figure 1.

The reason for this deviation becomes clear if we investigate the fraction of algorithms utilized by the agents at different memories. This is shown in figure 2 for the three selection procedures, and the four types of algorithms that may be utilized by the agents. We find that in all cases, the distributions converge once the agents extend their memory horizon past a critical value. That memory value also corresponds to the point where the scatter in the data in figure 1 disappears, and the data track the curves marked by the solid line. For scheme I, since the algorithms are freely and readily changed from the total pool, that position is rapidly reached at a short memory of $m = 4$. For schemes II and III, however, the reaching of the corresponding position occurs at $m = 9$ and the scatter in the data at lower memory values is reflected into figure 1. We therefore mark the approximate expected trace of the data, had they followed the converged distributions in all memory values, by the dashed line.

Also interesting is the fact that the point-wise hypotheses (cycle detectors) are the most frequently used algorithms in all schemes. In scheme I where the algorithms are changed immediately in case they fail, they appear nearly twice as much as all the other types, indicating that they are the more often winners at instantaneous steps. That they make-up ⅔ of the algorithms utilized in scheme II, where the agents change to the predictor with the least error, points to the fact that they are also more precise predictors. Furthermore, in scheme III which uses cumulative rewarding, the competitive edge of cycle detectors gets compounded, and 90% of the algorithms utilized are composed of these. This over-expression of the point-wise hypotheses hints that there are cycles that occur in the data, an observation that calls for further research in future studies. The remaining types behave almost equally well in scheme I, having values reduced slightly below that expected of random choice (23, 19, and 20% respectively for arithmetic average, weighted average, and trend, as opposed to the expected nearly 25% each); the reductions contribute almost equally to the enhancement in the cycle detectors. However, once rewarding is introduced in schemes II and III, weighted average algorithms are used less. In particular, in III their probability of appearance is nearly zero for all memories. Also, trend algorithms behave slightly better than arithmetic averages in II and III, opposite of that observed in scheme I.

In sum, for a given scheme, beyond a critical size of memory, the attendances fall onto an S-shaped curve. For shorter memories, attendances show large deviations from this curve. The critical memory where the average attendance begins to follow a predictable pattern corresponds to the converged distributions of the algorithm types used from the algorithm pool. The question remains, however, as to the origin of the bounded nature of attendance and its dependence on agent memory. We utilize a statistical mechanical approach to pose an answer to this question.

**A deductive descriptor of agent behavior in the El Farol bar**

In line with the findings of the previous section, we seek a potential that will be compatible with the general findings therein. We therefore use the following potential to describe the El Farol Bar system at a given point in time where the attendance is $a_o$:

$$E(\tau) = [L - a_o(\tau)]^2 + \alpha \sum_{i=1}^{N} \sum_{j=i+1}^{N} h_i(\tau) h_j(\tau) \qquad (3)$$

At time $\tau$, $h_i(\tau)$ is the preference of individual $i$ for attendance, as defined in equation 1. The first term drives the system towards the externally imposed constraint, $L$, the comfort level. The second term is for the interaction between individuals inside the system. The summation runs over all possible pairs of individuals, since the information is available to all individuals and there is no locality. $\alpha$ controls the strength between these two competing effects. The second term in equation 3 can be evaluated in terms of $a_o$ and $N$ so that:

$$E(\tau) = [L - a_o(\tau)]^2 + \frac{\alpha}{2}[(2a_o(\tau) - N)^2 - N] \qquad (4)$$

An ensemble describing the attendance to the bar may be generated by applying the usual Metropolis Monte Carlo scheme, where, at each step, $\tau$, a completely new attendance profile (in fact just a single value for $a_o$) is generated from a uniform distribution. We retain the previous attendance or accept the newly generated one according to the Boltzmann factor, exp [-$\beta$ ($E_{new}$ − $E_{old}$)]. $\beta$ is the equivalent of inverse temperature in physical systems.

Thus, an ensemble representing the states of the system can be reproduced and system properties of interest may be predicted without generating the individual preferences of the agents, and by taking the appropriate averages over that data. Alternatively, an analytical solution to the various system parameters may be obtained using the statistical mechanical treatment in the canonical ensemble, and assuming that the discrete nature of the system can be approximated by continuum if $N$ is large enough. The analytical solution for the average attendance, $A = <a_o>$, is then given by the expression:

$$A = \gamma + \frac{\exp(-b\gamma)}{a\sqrt{\pi}} \frac{1 - \exp[2Nb - (Na)^2]}{\text{erf}(a\gamma) + \text{erf}[a(N-\gamma)]}, \quad (5)$$

with the appropriate substitutions

$$a = \frac{\sqrt{(1+2\alpha)\beta}}{N}, \quad b = \frac{(L+\alpha N)\beta}{N^2}, \text{ and } \gamma = \frac{b}{a^2} = \frac{(L+\alpha N)}{(1+2\alpha)}. \quad (6)$$

For large $\beta$, the second term in equation 5 is negligible, and the average attendance approaches $\gamma$. Also, for $\alpha = 0$ in the limit $\beta \to \infty$, we recover the result obtained in many studies involving the agent based modeling of the bar problem: $A = L$.

With this model, we first we study the effect of $\beta$ and $\gamma$ on the system at a fixed value of the comfort level, $L = 60$. Results for three different values of $\alpha$ (0.0, 0.1, and 0.2) that describe the increasing contribution from the internal dynamics of the system relative to the externally imposed limitation are shown in figure 3. We find that attendance shows a transition in the range *ca.* 1 < $\beta$ < 10, for various values of $\alpha$. Note that the lower limit of $N/2$ is expected even when $\alpha = 0$, for small enough values of $\beta$. Hence, the first term in equation 3 is capable of describing the bounds on the attendance, [$N/2$, $L$]. However, it cannot describe the behavior observed in figure 1 for schemes II and III, that the system hits an upper bound that is lower than $L$.

**Unifying the inductive and deductive viewpoints of the bar problem**

A comparison of figures 1 and 3 indicates that the agent-based modeling and the statistical mechanical approach may be describing the same phenomena, whereby, agent memory $m$ is commensurate with the inverse temperature $\beta$, and the algorithm pool used (also influenced by the selection procedure) is reflected into the single parameter $\alpha$. Since figures 1 and 3 were reproduced for a single comfort level value, $L$, we reproduce the attendance profile of the agents for a large range of $L$s to validate the generality of equation 5. To achieve this, we first find the value of $\alpha$ at the particular $L$ as $\beta \to \infty$ (or, equivalently, as $m \to \infty$) that best describes the limiting value, $\gamma = (L + \alpha N) / (1 + 2\alpha)$, in the agent-based modeling (equation 6). We then select a particular memory $m$, and find the matching value of $\beta$, still at a single value of $L$. Finally, using this particular pair of ($\alpha$, $\gamma$), we reproduce the attendance profile of the agents for a range of values of $L$. In figure 4, we compare the agent-based and statistical mechanical modeling results for the three schemes I, II and III, studied in this work, for the comfort level range [30, 70]. As a demonstration, shortest memory cases that have equilibrated distributions of strategies are chosen for methods I, II, and III, ($m = 4$, 9, and 9, respectively, following the distributions in figure 2) as well as the long memory limit of method I. We find that the attendance profiles within a large range of values are well-reproduced by equation 5. Deviations are somewhat larger for method III, and this is due to the more impulsive change in the distribution of the algorithm types that survive with the selection strategy in this method, as the boundaries are approached for the comfort level (figure 4, bottom). For example, at this size of memory, and at $L = 60$, almost 90% of the agents use point-wise prediction strategies, and most of the remaining ones use linear fits to the available data (trend followers). On the other hand, by the point where $L = 80$, the distribution profiles are completely revamped with only half the agents using point-wise predictions, on average, and 45% using arithmetic averages, while all the remaining 5% are trend followers. Thus, the parameter $\alpha$ describes the distribution of strategies in the portfolios of the agents, independent of the memory, $m$. The latter is described by the parameter $\beta$.

**Concluding Remarks**

Whether the average attendance will converge to the externally provided comfort level or not depends on the algorithm selection procedures of the agents. Changing the algorithm used whenever it fails, irrespective of the past success of the algorithm, and picking up another one randomly (scheme I), drives the average attendance to the comfort level, $L$, as the agents use more information from the past. Taking into account a merit based stickiness to the algorithms employed in the past (schemes II and III), exhibit considerable deviation from the path that carries the average attendance to $L$. As shown in figure 1, stickiness not only alters the plateau levels, but also yields large fluctuations at the approach-to-plateau pathways, especially for shorter memory allocations (empty circles).

Information carrying capacity modulates the fractional use of the type of algorithms available in the pool. For short time horizons, it is rather difficult to estimate the distinct distributions. The algorithm clusters, namely, cycle detectors, trend followers, average takers, are shared by the agents sporadically, exhibiting a transient regime as observed in figure 2. As more of the past information is made available to all agents, independent of the rewarding scheme, fractional distributions are collectively balanced. Note, however, that reaching the detailed balance in the distributions does not coincide with reaching the plateau value of the attendance in figure 1, but is rather associated with the regions of small variations from the paths described by the S-shaped curves.

We fit a deductively constructed expression, the minimum of which gives the equilibrium distributions that are inductively obtained by the agents. This expression must take the limiting values of the problem as input: One limit is given by the comfort level, $L$, which is expected to be reached when perfect information is available to the agents; the other limit is $N/2$ which occurs when there is no information available to the agents and random behavior is observed. The competition between these two is moderated by a parameter $\alpha$. The average attendance is obtained analytically via standard canonical statistical mechanics calculations. The inverse temperature plays the role of the information allocated to the agents. It is then possible to associate a stickiness type, schemes I – III for instance, and the memory horizon $m$ from the agent-based calculations to the pair of competition moderating parameter $\alpha$ and the inverse temperature $\beta$.

The inductively organized S-shaped process is governed by the algorithm picking schemes while the deductively calculated one is controlled by tuning parameter $\alpha$ as displayed in figures 1 and 3, respectively. The equivalence between these two S-shaped processes is undertaken by different comfort levels as shown in figure 4, where the variables pair $\alpha$, $\beta$ predicted from a single value of $L$ is applicable to a wide range of comfort levels. A discrepancy between the analytically obtained average attendances and those calculated by agent-based simulations only emerges as the comfort level departs farther away from the symmetry, $N/2$. This departure can be traced to a modified sampling of the available algorithm pool by the agents (figure 2 and lower part of figure 4), and may be corrected by the choice of a modified variables pair $\alpha$, $\beta$.

The currently posed analytical model is capable of reproducing the *emergent* behavior that is observed in the collective actions of the agents in the El Farol bar problem. However, in its present state, the model cannot replicate the *adaptability* inherent in the behavior of the agents. This adaptability is exemplified by the higher than expected average attendance in the presence of limited information (low memory; figure 1), albeit with high fluctuations (data not shown), or the deviations from expectation for scheme III as the comfort level deviates considerably from $N/2$ (figure 4). Nevertheless, the current treatment paves the way for defining and exploring a plethora of new problems. Among these is the output of the system in the presence of variable threshold as a function of memory. It is also possible to include the effect of specific interactions between the agents at both the level of inductive and deductive models. One may also improve on the theory to take into account the presence of multiple equilibrium states available to the agents at a given value of the memory, so as to describe deviations of the attendance from that expected from the current form of the theory. Our general approach is to treat the findings from the agent-based modeling as experimental results obtained from the El Farol bar, and then use models to describe the main ingredients that govern the emergent and adaptive behavior patterns observed in this "laboratory."

**Acknowledgements.** We thank Güven Demirel and Ahmet Onur Durahim for valuable discussions. C.A. acknowledges support by the Turkish Academy of Sciences in the framework of the Young Scientist Award Program (CB/TÜBA-GEBIP/2004-3).

**REFERENCES**

1. *Adaptive Agents, Intelligence, and Emergent Human Organization: Capturing Complexity through Agent-Based Modeling* (2002) in *Proc. Natl. Acad. Sci. USA*, Vol 99 suppl. 3.
2. Tesfatsion L, Judd K, eds. *Handbook of Computational Economics*. Vol. 2. (2006).
3. Arthur WB (1994) *Amer. Econ. Rev.* 84: 406-411.
4. Arthur WB (1999) *Science* 284: 107-109.
5. Arthur WB (2006) in *Handbook of Computational Economics*, eds. L. Tesfatsion and K. Judd (North-Holland), ch. 32.
6. Challet D, Marsili M, Zhang Y-C (2005) *Minority Games: Interacting Agents in Financial Markets* (Oxford University Press).
7. Johnson NF, Jarvis S, Jonson R, Cheung P, Kwong YR, Hui PM (1998) *Physica A* 258: 230-236.
8. Challet D, Marsili M, Ottino G (2004) *Physica A* 332: 469-482.
9. Lus H, Aydin CO, Keten S, Unsal HI, Atilgan AR (2005) *Physica A* 346: 651-656.

**Figure Captions:**

**Figure 1.** Average attendance versus memory of agents for $N = 100$ agents, and a comfort level of $L = 60$. Each data point is the result of 50 runs of 2000 weeks. The first 100 weeks of attendance data are not included in the averages to remove transient effects. Error bars are smaller than the data points. Dashed lines approximate the paths that would have been followed at short memories, had the algorithm types followed the converged distribution profiles (see figure 2 and the text for details). Solid lines at longer memories are drawn through the points to guide the eye.

**Figure 2.** The distributions reached by each of the algorithm types (point-wise, arithmetic average, weighted average, and trend) at different memory values. Note that there are $4m - 3$ algorithms available to the agents at each memory value, $m$. The sum of the values at each $m$ is equal to one.

**Figure 3.** Average attendance versus $\beta$ from the analytical model for $N = 100$, and a comfort level of $L = 60$. Results are displayed for three different values of $\alpha$ that describes increasing relative contribution of the internal dynamics of the system to the externally imposed limitation. The Monte Carlo results, obtained from $10^6$ runs at each $\beta$, and the analytical solution (equation 5) give exactly matching results. The marked points on the curves represent the data points where the matching of parameters is made between the analytical and agent-based methodologies to reproduce the curves at a large range of comfort level values in figure 4 (blue dot is used for the blue curve, etc.).

**Figure 4. (Upper curve)** Predicted deviation of attendance from the analytical model of equation 5 (lines) compared to the results from the agent based modeling (data points), for several different cases of varying agent memory, $m$, and selection procedures (I, II, or III). Low memory cases with stable portfolio distributions are chosen for methods I, II, and III, ($m = 4$, 9, and 9, respectively, following the distributions in figure 2), as well as the long memory limit of method I. Matching the two parameters ($\alpha$, $\beta$) of the analytical solution at a single value of $L = 60$ (shown by the colored full circles; also see figure 3) faithfully reproduces the attendance profile for a wide range of $L$ [30, 70], except for method III, where the edge effects are more pronounced. **(Lower curve)** For the deviating (green) curve of method III, the distributions of algorithm types used by the agents are shown for the variety of comfort level values. Outside the range $37 - 63$ (marked by the dashed green lines) the distributions of utilized algorithms deviate significantly from those within the predictable range.

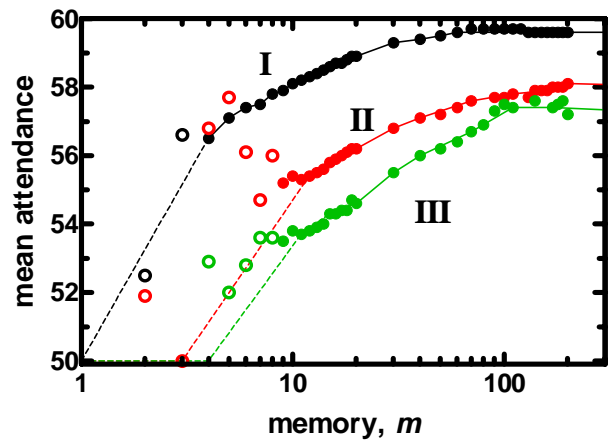

Figure 1

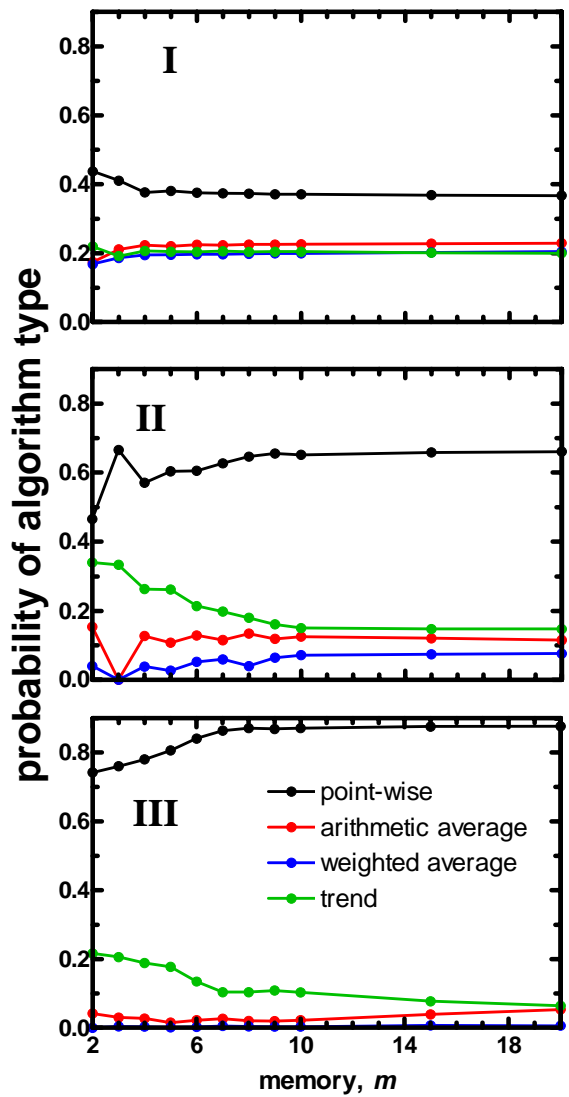

Figure 2

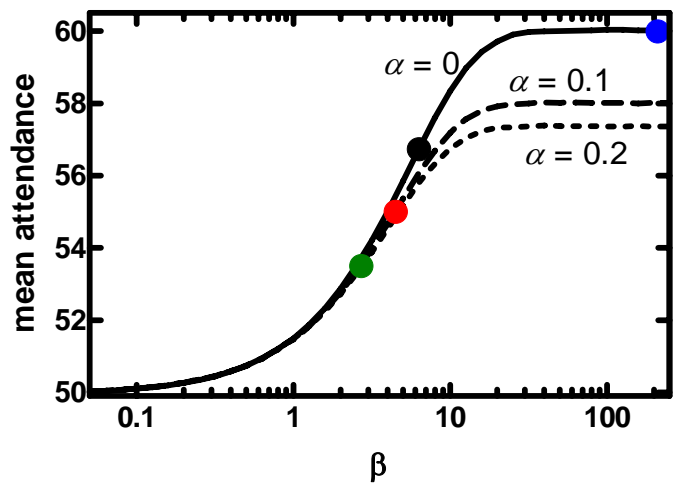

Figure 3

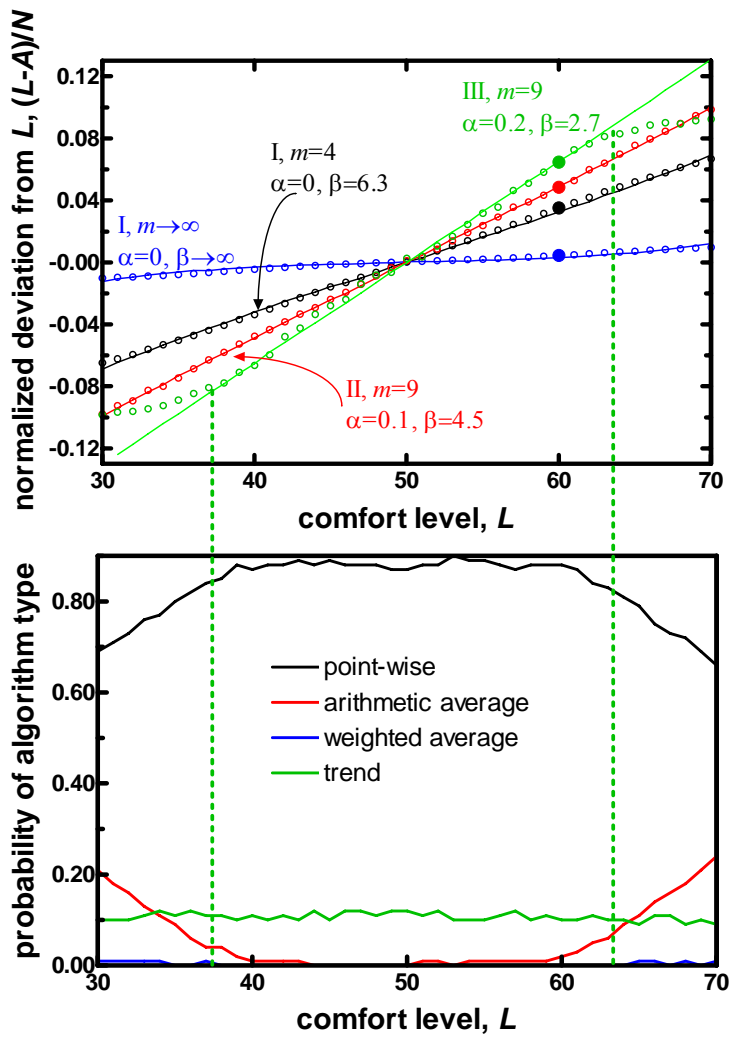

Figure 4